\documentclass[12pt]{iopart}
\begin{document}

\title[Coupled Yang-Mills Oscillators II]
      {Adventures of the Coupled Yang-Mills Oscillators:\\
       II. YM-Higgs Quantum Mechanics}

\author{Sergei G.~Matinyan$^1$ 
\footnote[1]{Present address: 3106 Hornbuckle Place, Durham, NC 27707, USA.}
and Berndt M\"uller$^2$}

\address{$^1$ Yerevan Physics Institute, 375036 Yerevan, Armenia}

\address{$^2$ Department of Physics, Duke University, Durham, NC 27708}

\date{\today}

\begin{abstract}
We continue our study of the quantum mechanical motion in the $x^2y^2$ 
potentials for $n=2,3$, which arise in the spatially 
homogeneous limit of the Yang-Mills (YM) equations. In the present
paper, we develop a new approach to the calculation of the partition
function $Z(t)$ beyond the Thomas-Fermi (TF) approximation by adding
a harmonic (Higgs) potential and taking the limit $v\to 0$, where $v$
is the vacuum expectation value of the Higgs field. Using the 
Wigner-Kirkwood method to calculate higher-order corrections in $\hbar$, 
we show that the limit $v\to 0$ leads to power-like singularities of 
the type $v^{-n}$, which reflect the possibility of escape of the 
particle along the channels in the classical limit. We show how these
singularities can be eliminated by taking into account the quantum 
fluctuations dictated by the form of the potential.
\end{abstract}

\maketitle

\section{Introduction}

We here continue our study of the quantum mechanical motion in the 
$x^2y^2$ potentials of phase space dimensions $2n$ with $n=2,3$, which 
arise in the spatially homogeneous limit of the Yang-Mills (YM) equations. 
As is well known \cite{bib1} (see \cite{bib2} for a review), these systems 
exhibit a rich chaotic behavior despite their extreme simplicity. 
Especially the $n=2$ model, the central object of this and also our
previous investigation (see \cite{bib*} - we will henceforth refer to
this work as I), has been widely studied.

Classically, this model possesses a logarithmically divergent volume of 
energetically accessible phase space $\Gamma_E$ \cite{bib2,bib3}
\footnote{This is in violation of Weil's famous theorem \cite{bib4}, which 
states that the average energy level density $dN/dE$ is asymptotically 
proportional to $\Gamma_E$.}, but its quantum mechanical version (YM 
quantum mechanics, YMQM) has a discrete spectrum \cite{bib5,bib6}.
Physically, the explanation is obvious: Quantum fluctuations, 
e.g.\ zero-point fluctuations, forbid that the trajector escapes along the 
$x$ or $y$ axis where the potential energy vanishes. The system is thus 
confined to a finite volume, and this implies the discreteness of the energy 
levels. Classically this escape is always possible without increasing energy. 
As we shall see below, these classically allowed configurations result in
singularities of the quasiclassical partition function.

In I we calculated the higher-order quantum corrections to the partition 
function (heat kernel) $Z(t)$ for the $x^2y^2$ potential using the
approximation \cite{bib7,bib8} based on the adiabatic separation of the
motion in $x$ and $y$ in the hyperbolic channels of the equipotential 
surface $xy={\rm const}$. The main assumption of this method is that
the final results of the calculations do not depend on the artificial 
boundary $Q$ dividing the central region $x,y\in [-Q,Q]$ from the 
channels $x,y\in [Q,\infty]$. We showed in I that this assumption,
after improvement of the quantum mechanical treatment of the motion in 
the channels, is correct not only for the Thomas-Fermi (TF) term but also 
for the leading (in powers of $tQ^4 \gg 1$) higher-order quantum corrections,
and we derived $Q$-independent asymptotic series in the parameter
$\lambda^2=g^2\hbar^4t^3$ for contribution of each region to $Z(t)$.

In the present paper, like in I, we explore the properties of the $x^2y^2$
model beyond the TF approximation, but we pursue a different approach. 
We calculate the higher-order corrections to $Z(t)$ as in \cite{bib9} 
by starting from the Yang-Mills-Higgs quantum mechanics (YMHQM) and 
taking the limit $v\to 0$, where $v$ is the vacuum expectation value
of the Higgs field. In the $n=2,3$ $x^2y^2$ models, $v$ determines the 
strength of the harmonic potential
\begin{eqnarray}
V(x,y) &= \frac{1}{2}v^2(x^2+y^2) &\qquad (n=2),
\nonumber \\
V(x,y,z) &= \frac{1}{2}v^2(x^2+y^2+z^2) &\qquad (n=3).
\nonumber
\end{eqnarray}
Due to the above mentioned logarithmic infinity of the classical phase
space volume of the $x^2y^2$ model it is impossible to completely 
disentangle the nonlinear coupled oscillators from the harmonic 
oscillations generated by the Higgs potential leading to a term 
in $Z(t)$ proportional to $\ln v$. In the $n=3$ case, which has a 
finite phase space volume at fixed energy, this method yields an
expression for $Z(t)$ that coincides in the limit $v\to 0$ with the
one obtained adiabatic separation method \cite{bib8}. This is
due to the negligible time spent by the classical trajectory in the 
depth of the hyperbolic channels \cite{bib8}. Higher-order corrections
change this situation essentially, as we shall see below.

Here we use the approach of ref.~\cite{bib9} with the limiting 
procedure $v\to 0$ for the calculation of $Z(t)$ beyond the TF 
approximation by applying the Wigner-Kirkwood (WK) method 
\cite{bib10,bib11,bib12} (see \cite{bib13} for a review of 
the WK method). The higher-order corrections ${\cal O}(\hbar^k)$ in the 
WK approach lead to a new phenomenon in the limit $v\to 0$: for 
$k\geq 2$ they yield power-like singularities of the form $v^{-k}$. 
These singularities are not cancelled at a given power $k$ as one 
might expect. The situation is completely different when one includes 
the quantum fluctuations in the channels of the $x^2y^2$ potential to 
all orders. These generate a confining potential, which does not disappear 
in the limit $v\to 0$, closes the flat direction, and eliminates the mentioned 
singularities, which are essentially classical. Taking these fluctuations 
into account, we are able to compare the expression for $Z(t)$ obtained 
by our method with the result for $Z(t)$ obtained by the method of 
\cite{bib7,bib8} in the TF approximation. 

Concerning the $n=3$ case with its finite phase-space volume, the 
singularities appear also at the higher-order corrections in contrast to 
the TF approximation and again are eliminated by the quantum fluctuations 
corresponding to the specific quartic form of the potential characteristic
of the YM quantum mechanics.\footnote{In the corresponding supersymmetric 
Yang-Mills quantum mechanical system, the wave function is not confined 
due to the cancellations between bosonic and fermionic degrees of freedom, 
and the spectrum is continuous \cite{bib19}.}

Finally, we develop a novel approximation scheme, which relies on a
resummation of certain terms in the KW expansion and thus introduces 
a nonvanishing Higgs potential, which avoids the divergences of the 
TF approximation. This approach is motivated by the need to take the 
quantum fluctuations inside the hyperbolic channels into account even 
in the lowest order approximation. We show that the new approach 
reproduces the TF result obtained with 
the method of \cite{bib7,bib8} without requiring an artificial subdivision 
of the phase space into different regions. A modified WK expansion can be
derived to systematically improve on this result.

In the next two sections we present the YMH system and the WK method of 
calculating $Z(t)$ beyond the TF approximation.

\section{Yang-Mills-Higgs classical and quantum mechanics: The
         Thomas-Fermi approximation}
 
There are several mechanisms that can suppress and even eliminate the 
classical chaos of the YM system (see \cite{bib2}). One is the Higgs 
mechanism \cite{bib14}. For spatially homogeneous fields (long wave 
length limit of YM system), if only the interaction of the gauge fields 
with the Higgs vacuum is considered, the classical Hamiltonian for $n=2$ 
is given by the expression 
\begin{equation}
H = \frac{1}{2}(p_x^2+p_y^2) + \frac{g^2}{2}x^2y^2 + \frac{v^2}{2}(x^2+y^2),
\label{eq01}
\end{equation}
where $v=\langle\phi\rangle$ is the vacuum expectation value of the 
Higgs field $\phi$. It is known \cite{bib14} that there is a classical 
transition from chaos to regular motion as $v$ gets large enough. More 
precisely, the chaos disappears when $g^2v^4/E > 0.6$ \cite{bib14}, 
where $E=H$ is the energy. The analogous transition in the adjacent 
energy-level spacing distribution was predicted \cite{bib15} and 
established in several papers \cite{bib16}. The quantized counterpart 
of (\ref{eq01}) is 
\begin{equation}
\hat H = -\frac{\hbar^2}{2}\nabla_{x,y}^2 + \frac{g^2}{2}x^2y^2 + 
          \frac{v^2}{2}(x^2+y^2).
\label{eq02}
\end{equation}
As in I, we measure all quantities in units of the energy $E$ with 
dimensions $[H]=1, [t]=-1, [x],[y]=1/4, [g]=0, [v]=1/4, [\hbar]=3/4$. 

It is obvious that the operator (\ref{eq02}) has a discrete spectrum as 
it has for $v=0$. The TF approximation to the heat kernel or partition 
function $Z(t) = {\rm Tr} [\exp(-t\hat H)]$ 
is the standard lowest-order semiclassical approximation valid for 
small $\hbar t^{3/4} \ll 1$. It is obtained by substituting the 
classical Hamiltonian for its quantum counterpart and replacing the 
trace of the heat kernel by the integral over the phase-space volume 
normalized by ($2\pi\hbar)^{-n}$, where $2n$ is the phase-space 
dimension. In other words, the TF approximation takes into account 
only the discreteness of the quantum mechanical phase space, but 
considers momenta and coordinates (in our case, the field amplitudes 
$x$ and $y$) as commuting variables. This method was used in numerous 
papers (see e.g. \cite{bib7,bib8,bib9}). For the calculation of the 
energy level density $\rho(E) = dN(E)/dE$ at asymptotic energies, the TF 
approximation is a consistent approach since, as we shall see below, all 
corrections to the TF term are structures with factors $\hbar^k t^\ell$ 
with $k,\ell$ positive integers. For the asymptotic energy level density 
$\rho(E)$ or $N(E)$ these corrections are negligible according the 
Karamata-Tauberian theorems \cite{bib5,bib6} relating the most singular 
part of $Z(t)$ to the asymptotic level density, 
$N(E) = \int dE \rho(E)= {\cal L}^{-1}[Z(t)/t]$, where ${\cal L}^{-1}$ 
denotes the inverse Laplace transform. 

In \cite{bib9} $Z(t)$ and $N(E)$ were calculated for the Hamiltonian 
(\ref{eq01}). We give the precise expression for $Z(t)$ of the YMHQM 
system in the TF approximation: 
\begin{equation}
Z(t) = \frac{1}{\sqrt{2\pi}g\hbar^2 t^{3/2}} 
       \exp\left(\frac{tv^4}{4g^2}\right) 
       K_0\left(\frac{tv^4}{4g^2}\right) ,
\label{eq03}
\end{equation}
where $K_0(z)$ is the modified Bessel function of the third kind. 
For the most interesting limit $v\to 0$ we get: 
\begin{equation}
Z(t) \to \frac{1}{\sqrt{2\pi}g\hbar^2 t^{3/2}} 
       \left(\ln\frac{8g^2}{tv^4} - C\right) ,
\label{eq04}
\end{equation}
where $C$ is the Euler constant. The impossibility of disentangling 
the coupled oscillators from the uncoupled ones is expressed by the 
logarithmic divergence of the phase space volume for $n=2$ as we 
already mentioned in the Introduction. Below we compare (\ref{eq04}) 
with the corresponding expression for $Z(t)$ obtained in \cite{bib7,bib8} 
for the pure $x^2y^2$ model. Because we shall often encounter the 
pre-factor appearing in (\ref{eq03}) and (\ref{eq04}) in the following, 
we introduce the special symbol $K$ for it:
\begin{equation}
K = (2\pi g^2\hbar^4 t^3)^{-1/2} \equiv (2\pi\lambda^2)^{-1/2} .
\label{eq04a}
\end{equation}

\section{Beyond the TF approximation: The Kirkwood-Wigner expansion}

In the present paper, unlike in I, we apply the WK expansion in all 
of phase space, avoiding the division of the phase space into a 
central region and hyperbolic channels. As we will see below, this
poses no problems as long as $v\neq 0$. However, singularities
appear in the limit $v\to 0$, unless the WK expansion is modified 
to include quantum fluctuations in a nonperturbative way.

Since we described the WK method in detail in I, we only give a very
brief outline here. We start from the equation (I--12), a set of recurrent
differential equations for the kernels $W_k$ of the partition function
$Z_k(t)$ at the $k$-th order in $\hbar$:
\begin{equation}
Z_k(t) = \frac{\hbar^k}{(2\pi\hbar)^n} \int d\Gamma\,
  W_k({\vec r},{\vec p};t)\, e^{-tH} ,
\label{eq11}
\end{equation}
where $d\Gamma = dxdydp_xdp_y$ and the classical Hamiltonian $H$ given
by (\ref{eq01}). We begin with the partition function at second order 
in $\hbar$ ($k=2$). Integrating over $p_x$ and $p_y$ and making use of 
the symmetry of the Hamiltonian (\ref{eq01}) with respect to the interchange 
$x \leftrightarrow y$, we obtain:
\begin{equation}
\int d\Gamma\, W_2\, e^{-tH}
= \frac{\pi t}{3}\left[ \left(-g^2 + \frac{tv^4}{2}\right)I_{10}
         + \frac{tg^4}{2}I_{21} - v^2I_{00} + tg^2v^2I_{11} \right] ,
\label{eq13}
\end{equation}
where we have used the notation (for $m\geq n$): 
\footnote{Note that the case $m=n$ needs to be calculated 
separately from the case $m>n$; see below.}
\begin{equation}
I_{mn} = 4 \int_0^\infty dx \int_0^\infty dy\, x^{2m} y^{2n}
  \exp\left[-\frac{t}{2}\left(v^2(x^2+y^2) + g^2x^2y^2\right) \right] .
\label{eq14}
\end{equation}
Note that the factors $\hbar^2$ from the KW expansion and from the
normalization of the phase space volume element have canceled, making
$Z_2$ independent of $\hbar$. Integrating over $y$ and introducing the 
new variable $u=g^2x^2/v^2$, we obtain: 
\begin{equation}
I_{mn} = \frac{\sqrt{2\pi}(2n-1)!!(v^2)^{m-n}}{t^{n+\frac{1}{2}}g^{2m+1}} 
         \int_0^\infty du\, u^{m-\frac{1}{2}}(1+u)^{-n-\frac{1}{2}}
         e^{-uz}
\label{eq14a}
\end{equation}
with $z=tv^4/2g^2$. The integral over $u$ is related to the Whittaker 
function $W_{\lambda,\mu}(z)$ (see \cite{bib17}, equation 9.222.1): 
\begin{equation}
I_{mn} = \frac{\sqrt{2\pi}(2n-1)!!(v^2)^{m-n}}{t^{n+\frac{1}{2}}g^{2m+1}} 
         e^{z/2} z^{-\frac{m-n+1}{2}} \Gamma\left(\frac{2m+1}{2}\right)
         W_{-\frac{m+n}{2},\frac{m-n}{2}}(z)
\label{eq15}
\end{equation}
For $m=n$ the Whittaker function has only a logarithmic divergence 
at $z=0$. For $m-n \geq 1$ power-like singularities appear. 
For completeness we give $Z_2(t)$ explicitly 
\begin{eqnarray}
Z_2(t) &=& \frac{tv^2}{12\sqrt{2\pi}gt^{1/2}}
           \left[ 2\left(-1+\frac{tv^4}{2g^2}\right)z^{-1}
           \Gamma\left(\frac{3}{2}\right) W_{-\frac{1}{2},\frac{1}{2}}(z) 
\right.
\nonumber \\
&& \qquad
         + z^{-1}\Gamma\left(\frac{5}{2}\right) 
           W_{-\frac{3}{2},\frac{1}{2}}(z) 
         - 2z^{-1/2}\Gamma\left(\frac{1}{2}\right) W_{0,0}(z)
\nonumber \\
&& \qquad \left.
         + 2z^{-1/2}\Gamma\left(\frac{3}{2}\right) W_{-1,0}(z) \right]
\label{eq16}
\end{eqnarray}
As is easily seen from (\ref{eq16}), $Z_2(t)$ has a singularity of the form
$v^{-2}$ at $v=0$. Using the limit of the Whittaker function
$W_{\lambda,\mu}(z)$ for small $z$, we find:
\begin{equation}
Z_2(t) \to -K \frac{\hbar^2g^2t^{3/2}}{6(tv^4)^{1/2}} \qquad (v\to 0)
\label{eq16a}
\end{equation}
For completeness we give here the $Z_2$ from the KW method in the 
limit $g=0$, i.~e.\ for two free harmonic oscillators using the 
asymptotic form of the Whittaker function: 
\begin{equation}
Z_2(t) = - \frac{1}{12} . 
\label{eq17}
\end{equation}
Together with the expression for the TF term \cite{bib9} we have 
\begin{equation}
Z_0(t) + Z_2(t) = \frac{1}{\hbar^2 v^2 t^2}
                  \left(1 - \frac{1}{12}\hbar^2v^2t^2\right) ,
\label{eq18}
\end{equation}
which are the first two terms in the Taylor expansion of the exact 
partition function for the two-dimensional harmonic oscillator at 
small $\hbar vt$:
\begin{equation}
Z(t) = \left[ 2\sinh\frac{1}{2}\hbar vt \right]^{-2} .
\label{eq18a}
\end{equation}

\section{Higher-order corrections and the limit $v\to 0$ for YMHQM}

At higher order $\hbar^k$ ($k\geq 2$) the mathematical structure of terms 
$W_k$ and $Z_k$ changes essentially. 
In the expression (\ref{eq14}) higher powers 
$m$ and $n$ appear and the difference between them increases ($m-n\geq 1$), 
causing the second index of the Whittaker function to exceed $\mu=1/2$. As 
a result, in the limit $z=tv^4/2g^2\to 0$ the finite sum of the power-like 
singular terms of $W_{\lambda,\mu}(z)$ begins to play a crucial role. 

A systematic analysis of the higher-order corrections using {\it Mathematica} 
leads to the conclusion that there is a correlation between the power of 
$\hbar$ (for $k\geq 2$) and the most singular terms in (\ref{eq15}): 
$m-n=k/2$ (due to the symmetry against exchange $x\leftrightarrow y$ 
it is always possible to put $m>n$). The case $m=n$ requires special 
consideration and, as found, leads only to logarithmic singularities.

It is easy to see that the next, less singular terms correspond to the case 
$m-n=(k/2)-2\ell$ ($\ell=1,2,\ldots; \ell < k/4$). For the general expression 
of $Z^{(m,n)}_k(t)$ we need to determine the powers of $g^2$ and $t$. For 
$g^2$ it is simply $g^{2m}$. To determine the power of $t$ we note that 
for each $k$ there are always terms without factors of $p_x$ and $p_y$ and 
minimal power of $t$ at given $k$. For such terms the factor is 
$t^{2m-n-1}$. For the terms containing factors of $p_x$ and $p_y$ the 
power of $t$ does not change after integration over the momenta. 
For the less singular terms with $m-n=\frac{1}{2}k-2\ell$ the corresponding
factor is $t^{2m-n-1-3\ell}$.

Now we are in position to write the general expression for the most singular 
terms contributing to the partition function at the order of $\hbar^k
(k=2,4,6,\ldots$). After integration over $p_x,p_y,x$ and $y$, keeping
only terms having $m-n=k/2$ ($k/2\leq m\leq k, 0\leq n\leq k/2$) we obtain: 
\begin{equation}
Z_k^{(m,n)}(t) = K \left( \frac{4g^4\hbar^4t^3}{v^4t} \right)^{k/4}
                   \Gamma\left(\frac{k}{2}\right) (2m-k-1)!! .
\label{eq38}
\end{equation}
We see that there are strong singularities at $v=0$, and further analysis 
shows that these are not cancelled by the summation of all most singular 
terms at a given $k$. For the less singular terms with $m-n=k/2-2\ell$
with $\ell<k/4$ we have:
\begin{equation}
Z_k^{(m,n,\ell}(t) = K \left( \frac{4g^4\hbar^4t^3}{v^4t} \right)^{k/4}
         \left(\frac{v^4t}{4g^4}\right)^\ell 
         \Gamma\left(\frac{k}{2}-\ell\right) (2m-k-1)!! .
\label{eq39}
\end{equation}
Due to the appearance of $(m-n)$ in the argument of the gamma function
in (\ref{eq39}), the case with $m=n$ must be considered separately. 
It is obvious that these terms have no power-like divergences, but only 
logarithmic singularities like the TF term. We obtain 
\begin{equation}
Z_k^{(m=n)}(t) = K (g^2\hbar^4t^3)^{k/4} \left[ \ln\frac{2g^2}{v^4t}
                   -2C - \psi\left(m+\frac{1}{2}\right) \right] ,
\label{eq40}
\end{equation}
where $\psi(x)$ is the logarithmic derivative of the Gamma function.
Let us briefly discuss these results. The power-like singularities in the KW 
approach for $k\geq 2$ are related to the possibility of escaping classically 
along the axis $x=0$ or $y=0$ where, in the limit $v\to 0$, the potential 
energy vanishes. Non-zero $v$ forbids such escape to infinity. These 
singularities affect any classically calculated distribution function, in 
particular, the heat kernel $Z(t)$. They also show that the trajectories lie 
deep inside the channels most of the time. Quantum mechanical fluctuations 
forbid any escapes along the axes. 

The $v^{-k}$ singularities have more resemblance with the infrared 
singularities, they are related to the behavior at long distances and 
different from the usual ultraviolet divergences connected with the 
asymptotic expansion in powers of $h^k$. The absence of power-like 
singularities for $m=n$ is easily explained: for $m=n$ the configurations 
dominate along the diagonals ($|x|=|y|$), whereas for $m\gg n$ the 
configurations populate the channels and have a trend to escape if 
quantum fluctuations do not forbid this. It is thus clear that we have 
to take into account the effect of the quantum fluctuations on the motion 
inside the channels, which the perturbative expansion of the WK method
fails to do.

\section{Quantum fluctuations and power-like singularities}

In this Section we will attempt to include quantum fluctuations created
by the form of $x^2y^2$ potential in a framework, which does not rely
directly on the adiabatic separation of the motion inside the hyperbolic 
channels as it was elaborated in detail in I. Let us consider the motion 
along the $x$-axis. The heat kernel for the $x^2y^2$ potential generates 
a mean spread in $y$ at the position $x$ of the order of $\delta y\sim 
(g^2x^2t/2)^{-1/2})$. Quantum mechanics dictates that the spread of the 
conjugate momentum $p_y$ is at least $\delta p_y \geq \hbar g|x|(t/2)^{1/2}$. 
Analogous relations hold between $y$ and $p_x$: $\delta p_x \geq 
\hbar g|y|(t/2)^{1/2}$ for the motion along the $y$-axis. We now propose a
modification of the WK formalism, which incorporates these relationships 
into the generating functional for the expansion in powers of $\hbar$.

To achieve this, we resum the term $-\frac{1}{2}\hbar^2 t(\Delta V)$ 
in the WK operator (I--10) to all orders by writing 
\begin{eqnarray}
W({\vec r},{\vec p};t) = {\tilde W}({\vec r},{\vec p};t) 
\exp\left( -\frac{\hbar^2}{4} t^2 \Delta V \right) .
\label{eq40a}
\end{eqnarray}
Inserting this definition into the differential equation (I--10) for the 
WK kernel $W$ yields an equation for the phase space function $\tilde W$:
\begin{eqnarray}
\frac{\partial \tilde W}{\partial t} 
& = & \frac{\hbar^2}{2}\left[ \Delta + t^2(\nabla V)^2 
  - \frac{2it}{\hbar}({\vec p}\cdot\nabla V) 
  - \frac{\hbar^2}{4} t^2 (\Delta\Delta V) \right.
\nonumber \\
& & \qquad
  + \frac{2}{\hbar}(i{\vec p} - \hbar t\nabla V)\cdot\nabla
  - \frac{\hbar^2}{2} t^2 \nabla(\Delta V)\cdot\nabla
\nonumber \\
& & \qquad \left. 
    - \frac{\hbar}{2} t^2 (i{\vec p} - \hbar t\nabla V)\cdot\nabla(\Delta V)
      \right] \tilde W .
\label{eq40b}
\end{eqnarray}
The term in the exponent of (\ref{eq40a}):
\begin{equation}
-\frac{\hbar^2}{4} t^2 \Delta V = - \frac{\hbar^2}{4} g^2 t^2 (x^2 + y^2)
\label{eq40c}
\end{equation}
acts like a Higgs potential with $v_{\rm eff}^2 = \hbar^2g^2t/2$. 
What is unusual about this term is that the effective potential itself 
is time-dependent. The connection to the argument given at the beginning
of this section becomes evident, when one recognizes that the exponent 
represents the lower bound associated with the kinetic energy demanded 
by the uncertainty relation:
\begin{equation}
\frac{1}{2} (\delta p_x^2 + \delta p_y^2) \geq \frac{\hbar^2}{4} t 
  \Delta V .
\label{eq40d}
\end{equation}

It is straightforward to derive a recursion relation analogous to (I--12)
for the coefficients of the expansion of $\tilde W$ in powers of $\hbar$:
\footnote{Note that the expansion of $\tilde W$ in powers of $\hbar$ does not
yield an expansion of $Z(t)$ strictly in powers of $\hbar$ because 
of the nonpolynomial factor containing $\hbar$ in (\ref{eq40a}).}
\begin{eqnarray}
\frac{\partial {\tilde W}_k}{\partial t} 
&=& i{\vec p}\cdot\left[\nabla - t(\nabla V)\right] {\tilde W}_{k-1} 
    + \frac{1}{2}\left[\Delta + t^2(\nabla V)^2 - 2t\nabla V\cdot\nabla\right] 
      {\tilde W}_{k-2} 
\nonumber \\
&& \qquad
    - \left[ \frac{it^2}{4} {\vec p}\cdot\nabla (\Delta V) \right]
      {\tilde W}_{k-3}
\nonumber \\
&& \qquad
    + \left[ \frac{t^3}{4}\nabla V\cdot\nabla(\Delta V) 
    - \frac{t^2}{4}\nabla(\Delta V)\cdot\nabla
    + \frac{t^2}{8}\Delta\Delta V \right] {\tilde W}_{k-4} .
\label{eq40d1}
\end{eqnarray}

Before we apply this trick to the quartic YM oscillator, it is useful
to briefly explore how it works for the harmonic oscillator. It is easy 
to see that (\ref{eq40d1}) yields the correct expansion for the partition 
function for the potential $V=\frac{1}{2}v^2x^2$. Indeed, by calculating 
${\tilde W}_0,{\tilde W}_2,{\tilde W}_4$ and integrating over $p$ and $x$, 
we obtain 
\begin{eqnarray}
{\tilde Z}_0(t) &=& \frac{1}{\hbar vt} e^{-(\hbar vt)^2/4} ,
\nonumber \\
{\tilde Z}_2(t) &=& \frac{5}{24} \hbar vt e^{-(\hbar vt)^2/4} ,
\nonumber \\
{\tilde Z}_4(t) &=& \frac{127}{5760} (\hbar vt)^3 e^{-(\hbar vt)^2/4} ,
\label{eq40d2}
\end{eqnarray}
giving the expansion in powers of $(\hbar vt)$ (up to $\hbar^4$) of
the exact partition function of the harmonic oscillator,
$Z(t) = (2\sinh\hbar vt/2)^{-1}$. Note that, in this case, $\tilde Z_k$
contributes to all powers in the $\hbar$ expansion of $Z(t)$ beginning 
with $\hbar^{k-1}$. 

Now consider the non-linear potential $\frac{1}{2}g^2x^2y^2$. 
The lowest term is easily obtained from (\ref{eq03}) by substituting 
$v \to v_{\rm eff}$:
\begin{equation}
{\tilde Z}_0 = K
  \exp\left( \frac{\hbar^4}{16}g^2t^3\right)
  K_0\left( \frac{\hbar^4}{16}g^2t^3\right) 
\approx K \left[ \ln\frac{1}{g^2\hbar^4 t^3} + 5\ln 2 - C \right] , 
\label{eq40e}
\end{equation}
which coincides (with logarithmic precision) with the result obtained
in \cite{bib7,bib8} for $Z(t)$. The second-order term is
\begin{equation}
{\tilde Z}_2 = \frac{g^2t}{24\pi}(4{\tilde I}_{10} + g^2t{\tilde I}_{21})
\label{eq40e1}
\end{equation}
with
\begin{equation}
{\tilde I}_{mn} = 4 \int_0^{\infty} dx \int_0^{\infty} dy\, x^{2m} y^{2n}
  \exp\left(-\frac{1}{2}g^2x^2y^2t - \frac{\hbar^2}{4}g^2(x^2+y^2)t^2\right) .
\label{eq40e2}
\end{equation}
These integrals correspond to those defined in (\ref{eq14}) with the
substitution $v^2 \to \hbar^2g^2t/2$. The exact analytical expression for 
these integrals in terms of Whittaker functions was given in (\ref{eq15}). 
In the following we retain only the first term in the (finite) power series 
expansion of the Whittaker function; later we shall correct for this
simplification. Using the condition $z = g^2\hbar^4t^3/8 \equiv \lambda^2/8
\ll 1$, allowing us to neglect terms involving $\ln\lambda^2$ compared with
terms of the form $\lambda^{-2}$, we obtain
\begin{eqnarray}
{\tilde I}_{10} = \frac{\sqrt{2\pi}}{(g^2t)^{1/2}} \frac{4}{(\hbar gt)^2} ,
\nonumber \\
{\tilde I}_{21} = \frac{\sqrt{2\pi}}{(g^2t)^{3/2}} \frac{4}{(\hbar gt)^2} ,
\label{eq40e3}
\end{eqnarray}
yielding the result
\begin{equation}
{\tilde Z}_2(t) = \frac{5}{3} K .
\label{eq40e4}
\end{equation}
For ${\tilde Z}_4$ we retain only terms with $m-n=2$ as the most singular 
ones in the limit $v\to 0$, based on the usual arguments. This gives: 
\begin{equation}
{\tilde Z}_4(t) = \frac{(\hbar gt)^4}{2\pi\hbar^2 t}
  \left[ \frac{1}{30}{\tilde I}_{20} + \frac{g^2t}{180}{\tilde I}_{31}
  + \frac{(g^2t)^2}{576}{\tilde I}_{42} \right] .
\label{eq40e5}
\end{equation}
For the needed integrals we obtain in the same approximation 
($\lambda^2 \ll 1$):
\begin{eqnarray}
{\tilde I}_{20} = \frac{\sqrt{2\pi}}{(g^2t)^{1/2}} \frac{16}{(\hbar gt)^4} ,
\nonumber \\
{\tilde I}_{31} = \frac{\sqrt{2\pi}}{(g^2t)^{3/2}} \frac{16}{(\hbar gt)^4} ,
\nonumber \\
{\tilde I}_{42} = \frac{\sqrt{2\pi}}{(g^2t)^{5/2}} \frac{48}{(\hbar gt)^4} .
\label{eq40e6}
\end{eqnarray}
Substituting (\ref{eq40e6}) into (\ref{eq40e5}) we obtain:
\begin{equation}
{\tilde Z}_4(t) = \frac{127}{180} K .
\label{eq40e7}
\end{equation}
Collecting all results, we have:
\begin{equation}
{\tilde Z}_{0+2+4}(t) = K \left[ \ln\frac{1}{g^2\hbar^4 t^3} 
  + 5\ln 2 - C + \frac{427}{180} \right] , 
\label{eq40f}
\end{equation}

We improve upon the approximation made above to the Whittaker function
and consider the contribution of all singular terms in the asymptotic
expansion of $W_{\kappa,\mu}(z)$ for small $z$:
\begin{equation}
\sum_{p=0}^{m-n-1} \Gamma(m-n-p)\Gamma\left(n+\frac{1}{2}+p\right)
                   \left(-\frac{\lambda^2}{8}\right)^p ,
\label{eq40f1}
\end{equation}
where again $\lambda^2 = g^2\hbar^4t^3$. The complete expression for the 
contribution to ${\tilde Z}_k$ ($k=2,4,6,\ldots$) from the most singular 
terms is
\begin{equation}
{\tilde Z}_k^{(m,n)} = K \frac{2^k (2n-1)!!}{\Gamma\left(n+\frac{1}{2}\right)}
   \sum_{p=0}^{\frac{1}{2}k-1} \Gamma\left(\frac{1}{2}k-p\right)
   \Gamma\left(n+\frac{1}{2}+p\right) \left(-\frac{\lambda^2}{8}\right)^p .
\label{eq40f2}
\end{equation}
For the less singular terms ($m-n=\frac{1}{2}k-2\ell$, $\ell<\frac{1}{4}k$)
we get
\begin{eqnarray}
{\tilde Z}_k^{(m,n,\ell)} &=& K \lambda^{2\ell} 
   \frac{2^{k-4\ell} (2n-1)!!}{\Gamma\left(n+\frac{1}{2}\right)}
\nonumber \\
   && \qquad
   \sum_{p=0}^{\frac{1}{2}k-2\ell-1} \Gamma\left(\frac{1}{2}k-2\ell-p\right)
   \Gamma\left(n+\frac{1}{2}+p\right) \left(-\frac{\lambda^2}{8}\right)^p ,
\label{eq40f3}
\end{eqnarray}
and for the logarithmic term ($m=n$) we obtain
\begin{equation}
{\tilde Z}_k^{(m=n)} = K \lambda^{\frac{1}{2}k}
   \left[-\ln\lambda^2 + 3\ln 2 - 2C - \psi\left(m+\frac{1}{2}\right)\right] .
\label{eq40f4}
\end{equation}
It is clear that these contributions again generate an asymptotic series 
in the small parameter $\lambda^2 = g^2\hbar^4t^3$ of the following form:
\begin{eqnarray}
Z(t) &=& K \left[ -\ln\lambda^2 + 5\ln 2 - C 
     + \sum_{k=2,4\ldots} 2^k \sum_{n=0}^{k/2} a_n^{(k/2)} 
     \frac{2^k (2n-1)!!}{\Gamma\left(n+\frac{1}{2}\right)} \right.
\nonumber \\
   && \qquad\left.
   \sum_{p=0}^{\frac{1}{2}k-1} \Gamma\left(\frac{1}{2}k-p\right)
   \Gamma\left(n+\frac{1}{2}+p\right) \left(-\frac{\lambda^2}{8}\right)^p 
   \right] ,
\label{eq40f5}
\end{eqnarray}
where the $a_n^{(k/2)}$ ($n=0,1,\ldots,\frac{k}{2}$) are the coefficients
of the structures $(g^2t)^n{\tilde I}_{mn}$ ($m-n=\frac{k}{2}$) in the 
expression $\int d\Gamma\,{\tilde W}_k e^{-Ht}$ ($k=2,4,6,\ldots$), analogous
to the coefficients $a_m^{(n)}$ ($m=0,1,\ldots,3n$) in (I--71). For $k=4$,
e.g., these numbers are:
\begin{equation}
a_0^{(2)} = \frac{1}{30} , \qquad
a_1^{(2)} = \frac{1}{180} , \qquad
a_2^{(2)} = \frac{1}{576} .
\label{eq40f6}
\end{equation}
We note that this asymptotic series closely resembles the one derived
in I using a completely different approach, splitting the $x-y$ plane
into two integration region and treating the quantum fluctuations 
exactly in the region containing the hyperbolic channels. The leading
logarithmic term is identical in both cases, but it is not clear that
the constant coincides. The expansion parameter $\lambda^2$ is the 
same for both series. Unfortunately, our inability to find a simple 
general algorithm for the coefficients, $a_n^{(k/2)}$ here and 
$a_m^{(n)}$ in I, has prevented us to compare the two results in detail.
The less singular terms also lead, after summation over $k$, to an 
asymptotic series in the parameter $\lambda^2$. The same is true for
the terms $Z_k^{(m,m)}$ with $m=n$.

At the end of Section 4 we commented on the analogy between the limit 
$v\to 0$ and the infrared problem. We now can make this analogy more
precise. The infrared limit corresponds to the behavior of the system
at large distances or deep inside the channels, in the terminology of 
the paper I. When we compare the limiting forms for $v\to 0$ of the 
quantities $I_{mn}$ derived here with the expression (I--25) for the 
analogous quantities derived for large distances ($tQ^4 \gg 1$) in I, 
we obtain the correspondence:
\begin{equation}
Q^{2(m-n)} \sim \frac{m-n}{\Gamma\left(n+\frac{1}{2}\right)}
                \frac{1}{(tv^2)^{m-n}} .
\label{eq41g}
\end{equation}
Evidently the limit $v\to 0$ corresponds to the limit $Q\to\infty$,
supporting our claim that it represents the infrared limit.

\section{The three dimensional YMHQM}

Finally, we briefly consider the $n=3$ case of the YMHQM model.
The Hamiltonian for $n=3$ YMH classical mechanics is: 
\begin{equation}
H = \frac{1}{2} (p_x^2 + p_y^2 + p_z^2) 
    + \frac{g^2}{2} (x^2y^2 + y^2z^2 + z^2x^2) .
\label{eq42}
\end{equation}
In its quantum counterpart, ${\vec p}\,^2$ is replaced with $-\hbar^2\nabla^2$.
As we know the contribution to $Z(t)$ from the channels is negligible in 
the TF approximation \cite{bib8} if we apply the condition $tQ^4\gg 1$.
This is due to the fact that deep in a channel, e.~g.\ along the $x$-axis, 
the energetically accessible phase space volume gets pinched as $x^{-2}$ 
and not as $x^{-1}$ as it for the $n=2$ Hamiltonian. This implies that the 
limit $v\to 0$ is smooth and the expressions for $Z(t)$ from \cite{bib8} 
and \cite{bib9} coincide at $v\to 0$. Already for the second correction to 
the TF term this is not true as we shall see below. In the WK approach for 
$Z_2$ using (I--12) and (\ref{eq11}) we have 
\begin{equation}
Z_2(t) = \frac{1}{(2\pi\hbar)^3} \int dp_x dp_y dp_z dx dy dz\,
  W_2({\vec p},{\vec x};t)\, e^{-Ht} ,
\label{eq43}
\end{equation}
with $W_2$ from (I--19). Integrating over the momenta, using the symmetry 
of the potential energy in (\ref{eq42}), we have 
\begin{equation}
Z_2(t) = \frac{t^{1/2}}{2(2\pi)^{3/2}\hbar} 
         \left[ -g^2I_1 + \frac{g^4t}{4}I_2 \right] ,
\label{eq44}
\end{equation}
where integrals $I_1$ and $I_2$ over $x,y,z$ are given by
\begin{eqnarray}
I_1 &=& \int_0^\infty dxdydz\, x^2 e^{-V(x,y,z)t} ;
\label{eq45a}
\\
I_2 &=& \int_0^\infty dxdydz\, x^2(y^2+z^2) e^{-V(x,y,z)t} ;
\label{eq45b}
\end{eqnarray}
with 
\begin{equation}
V(x,y,z) = \frac{g^2}{2} (x^2y^2 + y^2z^2 + z^2x^2) 
         + \frac{v^2}{2} (x^2 + y^2 + z^2) .
\label{eq46}
\end{equation}
Integrating over $x$ and using polar coordinates $r,\phi$ for the 
integrations over $y$ and $z$ we obtain: 
\begin{eqnarray}
I_1 &=& \frac{1}{2} \left(\frac{2\pi}{t}\right)^{3/2} \int_0^\infty r dr\, 
    \frac{\exp\left(-\frac{1}{16}tg^2r^4 - \frac{1}{2}tv^2r^2\right) 
     I_0\left(\frac{1}{16}tg^2r^4\right)}{(v^2 + g^2r^2)^{3/2}} ;
\label{eq46a}
\\
I_2 &=& \frac{1}{2} \left(\frac{2\pi}{t}\right)^{3/2} \int_0^\infty r^5 dr\, 
    \frac{\exp\left(-\frac{1}{16}tg^2r^4 - \frac{1}{2}tv^2r^2\right) 
    I_0\left(\frac{1}{16}tg^2r^4\right)}{(v^2 + g^2r^2)^{3/2}} ;
\label{eq46b}
\end{eqnarray}
where $I_0(z)$ denotes the modified Bessel function of the first kind. 
We see that $I_1$ is divergent at $r=0$ if $v=0$ in contrast to the TF term. 

Again, we obtain a smooth transition in the limit $v\to 0$ if we make the 
substitution used in Section V before taking the limit $v\to 0$. As for 
the $n=2$ model, here also the zero-point quantum fluctuations in the 
channels generate an effective Higgs potential 
$\frac{1}{4}\hbar^2g^2t(x^2+y^2+z^2)$ and render the integral $I_1$ 
convergent. After this substitution, introducing the new variable 
$u^2= \frac{1}{16}tg^2r^4$, we get:
\begin{equation}
Z_2(t) = \frac{\sqrt{2}t^{3/4}}{\hbar g^{1/2}}
         \left[ -J_0(\lambda) + 4 J_2(\lambda) \right]
\label{eq47}
\end{equation}
with $\lambda = gt^{3/2}\hbar^2 (\ll 1)$ and
\begin{equation}
J_b(\lambda) = \int_0^\infty u^b du\, 
    \frac{e^{-u^2-\lambda u}}{(\lambda+8u)^{3/2}} .
\label{eq48}
\end{equation}
Expanding the exponential function in the small parameter $\lambda$, the 
integral (\ref{eq48}) can be expressed as a finite sum of the generalized 
hypergeometric functions. Retaining the main terms we have 
\begin{eqnarray}
J_0(\lambda) \approx \frac{1}{4\sqrt{\lambda}} ,
\nonumber \\
J_2(\lambda) \approx \frac{\Gamma\left(\frac{3}{4}\right)}{32\sqrt{2}} .
\label{eq48a}
\end{eqnarray}
Thus, the second-order correction $Z_2(t)$ for the $n=3$ YM model becomes
\begin{equation}
Z_2(t) \approx L \left[ 1 - \frac{\Gamma\left(\frac{3}{4}\right)^3}
       {2^{5/4}\pi^{3/2}} (g^2\hbar^4 t^3)^{1/4} \right] ,
\label{eq49}
\end{equation}
where 
\begin{equation}
L = \frac{1}{2}\Gamma\left(\frac{1}{4}\right)^3 (2\pi^2g^2\hbar^4t^3)^{-3/4}
\label{eq50}
\end{equation} 
is the TF term found in \cite{bib8,bib9}. We see from (\ref{eq49}) that 
the quantum corrections at the order $\hbar^2$ are parametrically enhanced 
due to the quantum fluctuations generated by the effective Higgs potential 
$\frac{1}{2}\hbar^2g^2(x^2+ y^2+z^2)t$.

\section{Conclusions}

We have shown in I and here that the richness of the classical YM mechanics 
with the $x^2y^2$ potential translates into, and even gets amplified by,
the quantum mechanical properties of the system. The YM quantum 
mechanics exhibits a confinement property, which strongly influences
the quantum mechanical motion in the $x^2y^2$ potential. At higher
order in $\hbar$ (up to $\hbar^8$) this results in the vanishing of the
leading quantum corrections, when we correctly take into account this 
property for the motion in the hyperbolic channels. We convinced that 
this property survives at higher orders, although we did not explicitly 
demonstrate it.

Here we calculated the quantum corrections to the partition function
by adding a Higgs term to the potential, and found that power-like
singularities arise in the limit $v\to 0$. We associate these 
essentially classical singularities with the fact that the
Wigner-Kirkwood expansion does not take into account the effect 
of the quantum fluctuations on the motion within the channels.
When these fluctuations, which are dictated by the uncertainty
relation and the hyperbolic form of the channels, are taken into
account, the escape along the coordinate axes, which is classically 
allowed, is prohibited, and the singularities disappear. As a result,
the Thomas-Fermi term for the partition function acquires a 
renormalization expressed in terms of an asymptotic series in the 
parameter $\lambda^2=g^2\hbar^4t^3$ within both approaches. 

We hope that the lessons we elicited from the present study of the
higher-order quantum corrections to the homogeneous limit of the 
Yang-Mills equations will be useful for an improved understanding 
of the internal dynamics of the Yang-Mills quantum field theory.

\end{document}